\documentclass[twocolumn,english,reprint,superscriptaddress,aps,prd]{revtex4-2}
\usepackage[T1]{fontenc}
\usepackage[latin9]{inputenc}
\usepackage{babel}
\usepackage{float}
\usepackage{mathrsfs}
\usepackage{dsfont}
\usepackage{amsmath}
\usepackage{graphicx}

\makeatletter
\@ifundefined{textcolor}{}
{%
 \definecolor{BLACK}{gray}{0}
 \definecolor{WHITE}{gray}{1}
 \definecolor{RED}{rgb}{1,0,0}
 \definecolor{GREEN}{rgb}{0,1,0}
 \definecolor{BLUE}{rgb}{0,0,1}
 \definecolor{CYAN}{cmyk}{1,0,0,0}
 \definecolor{MAGENTA}{cmyk}{0,1,0,0}
 \definecolor{YELLOW}{cmyk}{0,0,1,0}
}

\usepackage{braket}
\usepackage{tikz}
\usetikzlibrary{quantikz2}

\makeatother

\begin{document}
\title{Generalized quantum measurement in spin-correlated hyperon-antihyperon decays}
\author{Sihao Wu}
\email[]{shwu@mail.ustc.edu.cn}
\affiliation{Department of Modern Physics, University of Science and Technology
of China, Hefei 230026, China}
\author{Chen Qian}
\email[]{qianchen@baqis.ac.cn}
\affiliation{Beijing Academy of Quantum Information Sciences, Beijing 100193, China}
\author{Yang-Guang Yang}
\affiliation{Institute of Modern Physics, Chinese Academy of Sciences, Lanzhou
730000, China}
\author{Qun Wang}
\affiliation{Department of Modern Physics, University of Science and Technology
of China, Hefei 230026, China}
\affiliation{School of Mechanics and Physics, Anhui University of Science and Technology, Huainan,Anhui 232001, China}

\begin{abstract}
The rapid developments of Quantum Information Science (QIS) have opened
up new avenues for exploring fundamental physics. Quantum nonlocality,
a key aspect for distinguishing quantum information from classical
one, has undergone extensive examinations in particles' decays through
the violation of Bell-type inequalities. Despite these advancements,
a comprehensive framework based on quantum information theory for
particle interaction is still lacking. Trying to close this gap, we
introduce a generalized quantum measurement description for decay processes
of spin-1/2 hyperons. We validate this approach by aligning it with
established theoretical calculations and apply it to the joint decay
of correlated $\Lambda\bar{\Lambda}$ pairs. We employ quantum simulation
to observe the violation of CHSH inequalities in hyperon decays. Our
generalized  measurement description is adaptable and can be extended to
a variety of high energy processes, including decays of vector mesons,
$J/\psi,\psi(2S)\rightarrow\Lambda\bar{\Lambda}$, in the Beijing
Spectrometer III (BESIII) experiment at the Beijing Electron Positron
Collider (BEPC). The methodology developed in this study can be applied
to quantum correlation and information processing in fundamental interactions.

\end{abstract}
\maketitle

\section{Introduction}

Quantum field theory based on quantum mechanics and special relativity
is an underlying theory for elementary particles and their interactions.
Quantum information theory can offer us a new perspective for the
study of elementary particles and their interactions~\citep{Afik:2022kwm,Altomonte:2023mug}.
An essential feature of quantum mechanics is quantum nonlocality characterized
by the violation of Bell-type inequalities~\citep{Bell:1964kc,Clauser:1974tg,Aspect:1981zz}.
Previously, these Bell-type inequalities have been widely applied
in photonic and atomic systems detecting quantum correlation of electromagnetic
interaction at low energy~\citep{Yin:2017ips,BIGBellTest:2018ebd}. High-energy
processes also provide an alternative testing ground for quantum nonlocality
in electroweak and strong interactions~\citep{Aspect2002,BESIII:2018cnd}
with increasingly precise data. Recently there are a lot of works
that have been done along this line in different particle systems,
e.g., the hyperon-antihyperon system in charmonium decays at Beijing
Electron Positron Collider (BEPC)~\citep{Tornqvist:1980af,Khan:2020seu,Qian:2020ini,Chen:2013epa,Shi:2019kjf,Perotti:2018wxm,Batozskaya:2023rek}, the $\Lambda$ hyperon pairs in string fragmentation ~\citep{Gong:2021bcp},
the top quark and anti-top quark ($t\bar{t}$) system produced at
Large Hadron Collider (LHC)~\citep{Aoude:2022imd,Afik:2022dgh,Fabbrichesi:2021npl,Afik:2022kwm,Afik:2020onf},
the leptons pairs from $e^{+}e^{-}$ annihilation or Higgs decay~\citep{Fabbrichesi:2022ovb},
and correlated vector bosons and Higgs bosons in high energy processes~\citep{Barr:2021zcp,Barr:2022wyq,Fabbrichesi:2023cev,Ashby-Pickering:2022umy}. 
It is worth mentioning that the quantum entanglement in string fragmentation 
in high energy processes such as electron-positron and electron-proton collisions 
can be studied using the spin correlation of the hyperon-antihyperon system~\citep{Gong:2021bcp,Barata:2023jgd}.   
The spin correlation of the hyperon-antihyperon system has also been studied 
in heavy-ion collisions and can provide information on 
the vorticity structure of the strong interaction matter~\citep{Pang:2016igs,Lv:2024uev}.  


However, there are still some difficulties in testing Bell-type inequalities
in particle physics. One difficulty is that it is hard to overcome
all loopholes especially in high energy interactions~\citep{Hao:2009kj}.
Another is that we would encounter different Bell-type inequalities
in different particle decays because decay parameters vary for different
particles, which seems to lack a general feature~\citep{Khan:2020seu,Shi:2019kjf,Qian:2020ini}.
For these reasons, we need to search for a general framework to describe
quantum properties in high energy processes with the help of quantum
information theory. Recently the decay processes of particles have
been characterized as quantum measurement processes~\citep{Khan:2020seu,Qian:2020ini}
with particle interactions interpreted as quantum channels~\citep{Altomonte:2023mug}.
Moreover, quantum tomography has been employed to analyze quantum
correlation in top-antitop quark pairs~\citep{Aoude:2022imd,Afik:2020onf,Afik:2022kwm}.
Despite these works, a general framework for particle interactions
in the viewpoint of quantum information theory has yet to be established.


The purpose of this paper is to explore a mapping between the generalized quantum measurement and the particle decay process. We will focus
on decays of spin-1/2 hyperons, especially $\Lambda\bar{\Lambda}$
hyperon pair from decays of spin-0 charmonia $\eta_{c}$ and $\chi_{c0}$.
We describe decay processes of spin-1/2 hyperons in the language of  generalized quantum measurement and quantum channel. The generalized  measurement
in Bloch-Fano representation is also introduced for decay processes. Then
we apply the method to study the quantum correlation in $\Lambda\bar{\Lambda}$
from decays of $\eta_{c}$ and $\chi_{c0}$. Finally, we perform the
quantum simulation to test the CHSH inequality in $\Lambda\bar{\Lambda}$
and their decay daughters on the quantum computing platform \emph{Quafu}. 


The paper is organized as follows. We briefly review the concept of  the generalized quantum measurement and decay parameters of spin-1/2 hyperons
in Sec.~\ref{sec:Preliminaries}, and then describe decay processes
of spin-1/2 hyperons in the framework of the generalized measurement
and quantum channel in Sec.~\ref{sec:Hyperon_Decay} and Sec.~\ref{sec:Joint_decay},
respectively. We perform the quantum simulation to test the CHSH inequality
in Sec.~\ref{sec:Quantum_simulation}. We have a discussion about
the connection between the quantum measurement and particle reaction
processes in Sec.~\ref{sec:Projective_me}. A summary of the main
result and outlook is given in Sec.~\ref{sec:discussions}.


\section{Preliminaries \label{sec:Preliminaries}}

In this section, we briefly review some concepts in quantum measurement
theory and hyperon decays, which will be used in the following sections.

\subsection{Generalized  measurement and quantum channel \label{subsec:general_mea}}

In Quantum Information Science~(QIS), the measurement postulate is
employed to describe the act of measurement on a quantum system~\citep{Nielsen2015}.
According to this postulate, the measurement processes in quantum
physics are subjected by a collection of \textit{measurement operators}
$\{M_{m}\}$. These processes are defined as \emph{generalized measurements},
distinguishing from the well-known \textit{projective measurements}
or \textit{von Neumann measurements}\textit{\emph{, which have to
be orthogonal.}} When we use the density operator $\rho$ to describe
a system being measured, the probability of obtaining a certain outcome
$m$ is given by 
\begin{equation}
\mathcal{P}(m)= \textrm{Tr}\left(M_{m}\rho M_{m}^{\dagger}\right),
\label{eq:pro}
\end{equation}
where $\{\mathcal{P}(m)\}$ are probability distributions for all
possible outcomes satisfying the non-negative condition $\mathcal{P}(m)\geq0$
and the normalization condition $\sum_{m}\mathcal{P}(m)=1$. These
two conditions lead to constraints on $\{M_{m}\}$: the positive semidefiniteness:
$M_{m}^{\dagger}M_{m}\geq0$, and the completeness: $\sum_{m}M_{m}^{\dagger}M_{m}=\mathds{1}$.
Based on the measurement postulate in quantum mechanics, the initial
state $\rho$ instantaneously transforms after the measurement to
the state $\rho_{m}$, 
\begin{equation}
\rho\mapsto\rho_{m}=\frac{M_{m}\rho M_{m}^{\dagger}}{\mathcal{P}(m)},\label{eq:post}
\end{equation}
where the subscript of $\rho_{m}$ corresponds to the outcome $m$.
One can also define the \textit{positive operator-valued measurement}~(POVM)~\citep{Nielsen2015}
through the generalized measurement as $F_{m}\equiv M_{m}^{\dagger}M_{m}$
with the probability $\mathcal{P}(m)=\textrm{Tr}(F_{m}\rho)$ following
Eq.~(\ref{eq:pro}). The POVM formalism has been used in some recent
works in high energy physics~\citep{Ashby-Pickering:2022umy}.


Nevertheless, sometimes we cannot accurately obtain measurement outcomes
in real experiments. That is, if we lose the track of some measurement
outcomes, the resulting quantum states are described by an ensemble
$\{\mathcal{P}(m),\rho_{m}\}$, which indicates that the post-measurement
state $\rho_{m}$ has the probability $\mathcal{P}(m)$. The post-measurement
state is then 
\begin{align}
\sum_{m}\mathcal{P}(m)\rho_{m} & =\sum_{m}\textrm{Tr}\left(M_{m}\rho M_{m}^{\dagger}\right)\frac{M_{m}\rho M_{m}^{\dagger}}{\textrm{Tr}\left(M_{m}\rho M_{m}^{\dagger}\right)}\nonumber \\
 & =\sum_{m}M_{m}\rho M_{m}^{\dagger},
 \label{eq:Kraus}
\end{align}
which can be taken as a quantum evolution generated by the measurement.
This process is often characterized as a \textit{quantum channel}
$\rho\mapsto\mathcal{E}(\rho)\equiv\sum_{m}M_{m}\rho M_{m}^{\dagger}$,
and the set of $\{M_{m}\}$ is called \emph{Kraus operators}. In this
paper, the measurement operators play the role of Kraus operators.


\subsection{Spin-1/2 particle as qubit \label{subsec:qubit}}

In the standard model of particle physics, matter particles (leptons
and quarks) are all spin-1/2 fermions. Baryons including protons and
neutrons that are made of quarks are also fermions. The ground states
of octet baryons ($n$, $p$, $\Sigma^{\pm}$, $\Sigma^{0}$, $\Lambda$,
$\Xi^{-}$, $\Xi^{0}$) are all spin-1/2 particles. When we focus
solely on the spin degree of freedom, we can map these spin-$1/2$
particles to the ``qubits'' which originate from QIS and refer to
two-level quantum systems. Then the spin state of the particle along
the spin quantization direction $z$ can be expressed by a qubit denoted
as $\left|0\right\rangle \equiv\left|\uparrow_{z}\right\rangle $
and $\left|1\right\rangle \equiv\left|\downarrow_{z}\right\rangle $.
In the context of QIS, the density operator describing a qubit can
be put in Bloch representation as
\begin{equation}
\rho=\frac{1}{2}\left(\mathds{1}+\boldsymbol{\sigma}\cdot\boldsymbol{s}\right),\label{eq:Bloch}
\end{equation}
where $\boldsymbol{\sigma}=(\sigma_{x},\sigma_{y},\sigma_{z})$ is
the vector of three Pauli matrices, $\mathds{1}$ denotes the $2\times2$
unity matrix, and the vector $\boldsymbol{s}$ is called \textit{Bloch
vector} or polarization vector to describe the polarization of the
qubit. The Bloch vector $\boldsymbol{s}$ can be obtained by $\boldsymbol{s}=\left\langle \boldsymbol{\sigma}\right\rangle =\textrm{Tr}(\rho\boldsymbol{\sigma})$.


\subsection{Decay width and parameters \label{subsec:decaywidth}}

Hyperons such as $\Lambda$, $\Sigma$, and $\Xi$ are heavier than
protons and neutrons and contain one or more strange quarks. In this
work, we focus on hyperon decays. A typical hyperon decay is to another
spin-1/2 baryon $B^{\prime}$ accompanied by a spin-0 meson $M$ denoted
as $B\rightarrow B^{\prime}M$. According to the effective Lagrangian,
the decay matrix element reads~\citep{ParticleDataGroup:2020ssz},
\begin{equation}
\mathcal{A}_{B\rightarrow B^{\prime}M}=G_{F}m_{M}^{2}\bar{u}_{B}(C_{1}-C_{2}\gamma_{5})u_{B^{\prime}},\label{eq:matrix_element}
\end{equation}
where $G_{F}$ is the Fermi constant, $m_{M}$ is the meson mass,
and $C_{1}$ and $C_{2}$ are parity-violating $S$-wave and parity-conserving
$P$-wave decay amplitudes. The partial decay width is given by 
\begin{equation}
\Gamma=\frac{G_{F}^{2}m_{M}^{4}|\boldsymbol{q}|}{4\pi m_{B}}\left(m_{B^{\prime}}+E_{B^{\prime}}\right)\left(|S|^{2}+|P|^{2}\right),\label{eq:decay_width}
\end{equation}
where $m_{B}$ and $m_{B^{\prime}}$ are the masses of the mother
and daughter baryons respectively, $\boldsymbol{q}$ and $E_{B^{\prime}}=\sqrt{|\boldsymbol{q}|^{2}+m_{B^{\prime}}^{2}}$
are the momentum and the energy of the daughter baryon in the rest
frame of the mother baryon respectively, and the $S$ and $P$-wave
amplitudes in Eq.~(\ref{eq:decay_width}) are connected with $C_{1}$
and $C_{2}$ in Eq.~(\ref{eq:matrix_element}) by $S=C_{1}$ and
$P=-|\boldsymbol{q}|C_{2}/(m_{B^{\prime}}+E_{B^{\prime}})$. These
$S$ and $P$ amplitudes are Lorentz scalars and are fixed for the
two-body decay. In the decay angular's distribution, we introduce
three real parameters~\citep{Lee:1957qs} 
\begin{equation}
\alpha=\frac{2\mathrm{Re}\left(S^{*}P\right)}{\left|S\right|^{2}+\left|P\right|^{2}},\;\beta=\frac{2\mathrm{Im}\left(S^{*}P\right)}{\left|S\right|^{2}+\left|P\right|^{2}},\;\gamma=\frac{\left|S\right|^{2}-\left|P\right|^{2}}{\left|S\right|^{2}+\left|P\right|^{2}},\label{eq:parameters}
\end{equation}
which satisfy $\alpha^{2}+\beta^{2}+\gamma^{2}=1$, so $\alpha$,
$\beta$ and $\gamma$ are all in the range $[-1,1]$. Note that one
can use another parameterization $\beta=\sqrt{1-\alpha^{2}}\sin\phi$
and $\gamma=\sqrt{1-\alpha^{2}}\cos\phi$ with $\phi\in(-\pi,\pi]$.


\section{Decay of spin-1/2 hyperons \label{sec:Hyperon_Decay}}

In this section, we present our description of hyperon decays based
on the theory of generalized quantum  measurement. We mainly focus on the
angular distribution of the daughter particle (baryon) and its connection
with the generalized measurement description.


\subsection{Angular distribution of daughter particle}

To study the angular distribution of the outgoing baryon, we can write
$\Gamma_{B\rightarrow B^{\prime}M}$ derived from Eq.~(\ref{eq:matrix_element})
in the form of an angular integration 
\begin{equation}
\Gamma\propto\int\sum_{i=\pm}\left|\chi_{B^{\prime}}^{i}(S+P\boldsymbol{\sigma}\cdot\boldsymbol{n})\chi_{B}\right|^{2} \, d\Omega,
\label{eq:width_formula}
\end{equation}
where $\chi_{B}$ and $\chi_{B^{\prime}}$ represent spinors of mother
and daughter baryons respectively, and $\boldsymbol{n}\equiv\boldsymbol{q}/|\boldsymbol{q}|$
represents the momentum direction of $B^{\prime}$. The summation
indicates the average over daughter baryon's spin state. We can rewrite
Eq.~(\ref{eq:width_formula}) in a new form as 
\begin{equation}
\frac{1}{\Gamma}\frac{d\Gamma}{d\Omega}=\textrm{Tr}\left(M_{\boldsymbol{n}}\rho_{B}M_{\boldsymbol{n}}^{\dagger}\right),\label{eq:angular_dis}
\end{equation}
where $\rho_{B}$ denotes the spin density matrix for the mother baryon
$B$, $d\Omega=d\boldsymbol{n}=d\cos\theta d\phi$, and the operator
$M_{\boldsymbol{n}}$ is defined as 
\begin{equation}
M_{\boldsymbol{n}}\equiv\frac{1}{\sqrt{4\pi\left(|S|^{2}+|P|^{2}\right)}}\left(S+P\boldsymbol{\sigma}\cdot\boldsymbol{n}\right).\label{eq:M_n}
\end{equation}
We see in Eq.~(\ref{eq:angular_dis}) that the determination of the
angular distribution implies a generalized measurement characterized by
the measurement operator $M_{\boldsymbol{n}}$. In this way, the particle
decay can be regarded as a kind of generalized measurement process. The
only difference here is that the discrete measurement outcome $m$
described in Sec.~\ref{subsec:general_mea} to a continuous direction
$\boldsymbol{n}$ for the outgoing daughter baryon in experiments.
It is necessary to validate the positive semidefiniteness and completeness
of $\{M_{\boldsymbol{n}}\}$ as 
\begin{align}
M_{\boldsymbol{n}}^{\dagger}M_{\boldsymbol{n}}= \frac{1}{4\pi}\left(\mathds{1} + \alpha\boldsymbol{\sigma}  \cdot\boldsymbol{n}\right) & \geq  0, \ \ \mathrm{for}\ |\alpha|\leq1 \nonumber \\
\int M_{\boldsymbol{n}}^{\dagger}M_{\boldsymbol{n}} \, d\Omega= & \mathds{1}.
\end{align}
These two criteria ensure that $\{M_{\boldsymbol{n}}\}$ are legitimate
measurement operators in QIS.


After defining the measurement operators, we proceed to calculate
the decay process in this approach. According to the quantum measurement
postulate, the probability for the momentum direction of the daughter
baryon along $\boldsymbol{n}$ is given by $\mathcal{P}(\boldsymbol{n})=\textrm{Tr}(M_{\boldsymbol{n}}\rho_{B}M_{\boldsymbol{n}}^{\dagger})$,
which exactly equals to $(1/\Gamma)d\Gamma/d\Omega$ in Eq.~(\ref{eq:angular_dis}).
Given the initial spin density operator 
\begin{equation}
\rho_{B}=\frac{1}{2}(\mathds{1}+\boldsymbol{\sigma}\cdot\boldsymbol{s}_{B}),\label{eq:rho-b}
\end{equation}
as in Eq.~(\ref{eq:Bloch}), the resulting probability or the angular
distribution reads 
\begin{equation}
\mathcal{P}(\boldsymbol{n})=\textrm{Tr}(M_{\boldsymbol{n}}\rho_{B}M_{\boldsymbol{n}}^{\dagger})=\frac{1}{\Gamma}\frac{d\Gamma_{B\rightarrow B^{\prime}M}}{d\Omega}=\frac{1}{4\pi}\left(1+\alpha\boldsymbol{s}_{B}\cdot\boldsymbol{n}\right),
\label{eq:angular_re}
\end{equation}
which was first derived in Ref.~\citep{Cronin:1963zb}. 


Now we look at the polarization of the daughter baryon. According
to the quantum measurement postulate, the spin density operator of
the daughter baryon can be obtained via the post-measurement state
in Eq.~(\ref{eq:post}) as 
\begin{align}
\rho_{B^{\prime}}(\boldsymbol{n}) & =\frac{M_{\boldsymbol{n}}\rho_{B}M_{\boldsymbol{n}}^{\dagger}}{\textrm{Tr}\left(M_{\boldsymbol{n}}\rho_{B}M_{\boldsymbol{n}}^{\dagger}\right)}\nonumber \\
 & =\frac{1}{2}\left(\mathds{1}+\boldsymbol{\sigma}\cdot\boldsymbol{s}_{B^{\prime}}\right),\label{eq:post_Bloch}
\end{align}
where $\boldsymbol{s}_{B^{\prime}}$ is the polarization vector of
the daughter baryon in the mother baryon's rest frame as a function
of $\boldsymbol{n}$. By substituting $\rho_{B}$ in Eq. (\ref{eq:rho-b})
and $M_{\boldsymbol{n}}$ in Eq.~(\ref{eq:M_n}) into Eq.~(\ref{eq:post_Bloch}),
we obtain 
\begin{equation}
\boldsymbol{s}_{B^{\prime}}=\frac{(\alpha+\boldsymbol{s}_{B}\cdot\boldsymbol{n})\boldsymbol{n}+\beta(\boldsymbol{s}_{B} \times \boldsymbol{n})+\gamma\boldsymbol{n}\times(\boldsymbol{s}_{B}\times\boldsymbol{n})}{1+\alpha\boldsymbol{s}_{B}\cdot\boldsymbol{n}},\label{eq:Bloch_dau}
\end{equation}
where parameters $\alpha$, $\beta$ and $\gamma$ are defined in
Eq.~(\ref{eq:parameters}). The explicit expression of $\boldsymbol{s}_{B^{\prime}}$
in Eq.~(\ref{eq:Bloch_dau}) was initially derived by Lee and Yang~\citep{Lee:1957qs},
but here we rederived it in the language of  generalized quantum measurement.


We can look at the decay process as a quantum channel induced by $\{M_{\boldsymbol{n}}\}$.
In the hyperon decay process, the daughter baryon may fly in any direction
$\boldsymbol{n}$ associated with the probability $\mathcal{P}(\boldsymbol{n})$,
which is just the angular distribution of the daughter baryon in Eq.~(\ref{eq:angular_re})
that can be detected in particle physics experiments. With the daughter's
spin density operator $\rho_{B^{\prime}}(\boldsymbol{n})$, we have
an ensemble $\{\mathcal{P}(\boldsymbol{n}),\rho_{B^{\prime}}(\boldsymbol{n})\}$.
According to the generalized measurement postulate in Eq.~(\ref{eq:Kraus}),
this post-measurement ensemble can be interpreted as a quantum evolution
in the channel $\mathcal{E}$ defined as
\begin{align}
\mathcal{E}(\rho_{B})= & \int d\Omega \, M_{\boldsymbol{n}}\rho_{B}M_{\boldsymbol{n}}^{\dagger}\equiv\int d\Omega \, \mathcal{E}_{\boldsymbol{n}}(\rho_{B})\nonumber \\
= & \frac{1}{2}\left(\mathds{1}+\frac{1+2\gamma}{3}\boldsymbol{\sigma}\cdot\boldsymbol{s}_{B}\right),
\label{eq:averaged_Bloch}
\end{align}
which is actually the ensemble average of the daughter baryon's spin
density operator $\overline{\rho}_{B^{\prime}}=\mathcal{E}(\rho_{B})$,
and the term $(1+2\gamma)\boldsymbol{s}_{B}/3$ represents the average
polarization of the daughter baryon in the rest frame of the mother
hyperon.


In summary, we have established in this subsection the quantum measurement
interpretation for the nonleptonic decay of spin-1/2 hyperons. In
the next subsection, we will introduce an alternative representation
for the decay process.


\subsection{Bloch-Fano representation }

As we mentioned in subsec.~\ref{subsec:qubit}, a single qubit can
be expressed in a Bloch form as $\rho=(1/2)\sum_{\mu=0}^{3}r_{\mu}\sigma_{\mu}$
with $\sigma_{0}=\mathds{1}$ and $r_{\mu}\equiv[1|\boldsymbol{r}]^{\mathrm{T}}$.
Here the four coefficients $r_{\mu}$ has been put into a $4\times1$
column vector. For the initial spin density operator $\rho_{B}$,
we have $s_{B\mu}=[1|\boldsymbol{s}_{B}]^{\mathrm{T}}$ following Eq. (\ref{eq:rho-b}).
For the unnormalized density operator $\tilde{\rho}_{B^{\prime}}$
resulting from the measurement $\tilde{\rho}_{B^{\prime}}=M_{\boldsymbol{n}}\rho_{B}M_{\boldsymbol{n}}^{\dagger}$,
it can also be expressed in the Bloch representation as $\tilde{\rho}_{B^{\prime}}=(1/2)\sum_{\mu=0}^{3}\tilde{r}_{\mu}\sigma_{\mu}$
with $\tilde{r}_{\mu}\equiv[\tilde{r}_{0}|\tilde{\boldsymbol{r}}]^{\mathrm{T}}$.
As a consequence, the act of $M_{\boldsymbol{n}}$ can be interpreted
as a mapping $\mathcal{E}_{\boldsymbol{n}}(\rho_{B})$ that transforms
$s_{B}^{\mu}$ to $\tilde{r}_{\mu}$. 


Without the loss of generality, we assume $\boldsymbol{n}=\hat{\boldsymbol{z}}=(0,0,1)^{\textrm{T}}$.
According to Eq.~(\ref{eq:post_Bloch}) and Eq.~(\ref{eq:Bloch_dau}),
the mapping $\mathcal{E}_{\boldsymbol{n}}$ reads 
\begin{align}
1 & \mapsto\tilde{r}_{0}=\frac{1}{4\pi}\left(1+\alpha\hat{\boldsymbol{z}}\cdot\boldsymbol{s}_{B}\right),\nonumber \\
\boldsymbol{s}_{B} & \mapsto\tilde{\boldsymbol{r}}=\frac{1}{4\pi}\left(\mathcal{O}_{z}\boldsymbol{s}_{B}+\alpha\hat{\boldsymbol{z}}\right),
\end{align}
where $\mathcal{O}_{z}$ is a $3\times3$ matrix 
\begin{equation}
\mathcal{O}_{z}=\begin{bmatrix}\gamma & \beta & 0\\
-\beta & \gamma & 0\\
0 & 0 & 1
\end{bmatrix},
\end{equation}
As a result, the measurement process $\tilde{\rho}_{B^{\prime}}=M_{z}\rho_{B}M_{z}^{\dagger}=\mathcal{E}_{z}(\rho_{B})$
can be expressed in the matrix form, 
\begin{align}
\left[\begin{array}{c}
\tilde{r}_{0}\\
\hline \tilde{\boldsymbol{r}}
\end{array}\right] & =\mathcal{M}_{z}\left[\begin{array}{c}
1\\
\hline \boldsymbol{s}_{B}
\end{array}\right]\nonumber \\
\mathcal{M}_{z}\equiv & \frac{1}{4\pi}\left[\begin{array}{c|c}
1 & \alpha\hat{\boldsymbol{z}}^{\text{T}}\\
\hline \alpha\hat{\boldsymbol{z}} & \mathcal{O}_{z}
\end{array}\right]=\frac{1}{4\pi}\left[\begin{array}{c|ccc}
1 & 0 & 0 & \alpha\\
\hline 0 & \gamma & \beta & 0\\
0 & -\beta & \gamma & 0\\
\alpha & 0 & 0 & 1
\end{array}\right]\left[\begin{array}{c}
1\\
\hline s_{B}^{1}\\
s_{B}^{2}\\
s_{B}^{3}
\end{array}\right].\label{eq:Fano_z}
\end{align}
For an arbitrary direction $\boldsymbol{n}(\theta,\phi)$ in the quantum
measurement, we can use a SO(3) rotation matrix $\mathcal{R}(\phi,\theta,0)$
to rotate $\hat{\boldsymbol{z}}$ to $\boldsymbol{n}$ as $\boldsymbol{n}=\mathcal{R}\hat{\boldsymbol{z}}$.
Thus $M_{\boldsymbol{n}}$ can be put into a $4\times4$ matrix form
\begin{equation}
\mathcal{M}_{\boldsymbol{n}}=\frac{1}{4\pi}\left[\begin{array}{c|c}
1 & \alpha\boldsymbol{n}^{\text{T}}\\
\hline \alpha\boldsymbol{n} & \mathcal{O}_{\boldsymbol{n}}
\end{array}\right]=\left[\begin{array}{c|c}
1 & \boldsymbol{0}^{\textrm{T}}\\
\hline \boldsymbol{0} & \mathcal{R}
\end{array}\right]\mathcal{M}_{z}\left[\begin{array}{c|c}
1 & \boldsymbol{0}^{\textrm{T}}\\
\hline \boldsymbol{0} & \mathcal{R}^{-1}
\end{array}\right],\label{eq:Fano_n}
\end{equation}
which is merely a similarity transformation on the $4\times4$ matrix
$\mathcal{M}_{z}$. 


We can see in Eq.~(\ref{eq:Fano_n}) that all information about the
generalized measurement is encoded in the $4\times4$ matrix $\mathcal{M}_{\boldsymbol{n}}$.
The representation in Eq.~(\ref{eq:Fano_z}) and Eq.~(\ref{eq:Fano_n})
are called \textit{Bloch-Fano representation}~\citep{Fano:1957zz,Fano:1983zz,Benenti2011}
in QIS, which provides an alternative representation for the quantum
measurement. We should note that there is no fundamental distinction
between Eq.~(\ref{eq:post_Bloch}) and Bloch-Fano representation
in Eqs.~(\ref{eq:Fano_z}, \ref{eq:Fano_n}). Moreover, the
post-measurement ensemble $\mathcal{E}(\rho_{B})$ also has a Bloch-Fano
representation, which can be directly obtained from Eq.~(\ref{eq:averaged_Bloch})
as 
\begin{equation}
\mathcal{E}(\rho_{B})\Leftrightarrow\mathcal{M}=\left[\begin{array}{c|c}
1 & \boldsymbol{0}^{\textrm{T}}\\
\hline \boldsymbol{0} & \frac{1+2\gamma}{3}\mathds{1}_{3}
\end{array}\right].
\end{equation}


Note that the Fano matrix $\mathcal{M}_{z}$ in Eq.~(\ref{eq:Fano_z})
is exactly the same as \textit{aligned decay matrices} in Ref.~\citep{Perotti:2018wxm,Batozskaya:2023rek},
where the authors derived decay matrices through the helicity amplitude
method introduced by Jacob and Wick~\citep{Jacob:1959at}. This consistency
demonstrates the validity of the quantum measurement interpretation
in particle decay processes. It is possible to extract Fano matrices
in various decays of spin-1/2 hyperons \citep{Batozskaya:2023rek}, e.g.,
$B\rightarrow B^{\prime}\ell^{-}\bar{\nu}_{\ell}$, $B\rightarrow B^{\prime}\gamma$,
$B\rightarrow\ell^{+}\ell^{-}$ and $B\rightarrow B^{\prime}\pi^{+}\pi^{-}$.
We should emphasize that $B$ and $B^{\prime}$ in this paper represent
baryons, not B-meosns. 


\subsection{Decay chains}

It is common for the daughter baryon to undergo subsequent decay,
for example, in the decay chain $B\rightarrow B_{1}M_{1}\rightarrow B_{2}M_{2}M_{1}$,
in which $B_{1}$ decays to $B_{2}M_{2}$. This cascading decay process
can be described by \textit{the concatenate quantum measurement}.
Then the joint angular distribution or joint probability is given
by 
\begin{align}
\mathcal{P}(\boldsymbol{n}_{1},\boldsymbol{n}_{2}) & =\frac{1}{\Gamma}\frac{d\Gamma}{d\Omega_{1}d\Omega_{2}}\nonumber \\
 & =\textrm{Tr}\left[M_{\boldsymbol{n}2}^{1 \to 2}M_{\boldsymbol{n}1}^{B\to 1}\rho_{B}M_{\boldsymbol{n}1}^{\dagger B\to 1}M_{\boldsymbol{n}2}^{\dagger 1 \to 2}\right],
 \label{eq:decay_chain}
\end{align}
which is the probability that $B_{1}$ moving in the direction $\boldsymbol{n}_{1}$
decays to $B_{3}$ moving in the direction $\boldsymbol{n}_{2}$.
Here $\{M_{\boldsymbol{n}1}^{B\to 1}\}$ and $\{M_{\boldsymbol{n}2}^{1 \to 2}\}$
are two sets of measurement operators characterizing two decays, $B\rightarrow B_{1}M_{1}$
and $B_{1}\rightarrow B_{2}M_{2}$, respectively. 


Likewise, the decay chain can also be presented in the Bloch-Fano\textbf{
}representation as 
\begin{equation}
\mathcal{E}_{\boldsymbol{n}2}\circ\mathcal{E}_{\boldsymbol{n}1}(\rho_{B})=\mathcal{M}_{\boldsymbol{n}2}\mathcal{M}_{\boldsymbol{n}1}s_{B\mu},
\end{equation}
where $\mathcal{M}_{\boldsymbol{n}1}$ and $\mathcal{M}_{\boldsymbol{n}2}$
are Fano matrices in the form of Eq.~(\ref{eq:Fano_n}). Similarly,
the concatenate quantum measurement in Eq.~(\ref{eq:decay_chain})
can be extended to longer decay chains. 


\section{Joint decay of $B\bar{B}$ and two-qubit correlations \label{sec:Joint_decay}}

In this section, we will give a generalized quantum measurement introduced
in Sec.~\ref{sec:Hyperon_Decay} for the joint decay of $B\bar{B}$
to $B^{\prime}\bar{B}^{\prime}$ and study the correlation between
$B\bar{B}$ and $B^{\prime}\bar{B}^{\prime}$. 

If we only consider the spin degree of freedom for two spin-1/2 particles,
their joint state can be written in a general form 
\begin{align}
\rho_{B\bar{B}}=\frac{1}{4}\biggl(  \mathds{1}_{4}+\boldsymbol{s}_{B}\cdot\boldsymbol{\sigma}\otimes\mathds{1}_{2} & +\mathds{1}_{2}\otimes\boldsymbol{s}_{\bar{B}}\cdot\boldsymbol{\sigma} \nonumber \\ 
& + \sum_{i,j}C_{ij}\sigma_{i}\otimes\sigma_{j}\biggr),
\label{eq:two_qubit}
\end{align}
where $i,j=1,2,3$, $\mathds{1}_{n}$ is the $n\times n$ unity matrix,
$\boldsymbol{s}_{B}$ and $\boldsymbol{s}_{\bar{B}}$ are spin polarization
vectors for $B$ and $\bar{B}$ respectively, and $C_{ij}$ is a $3\times3$
real matrix for the spin correlation between $B$ and $\bar{B}$.
So there are 15 real parameters in $\rho_{B\bar{B}}$ corresponding
to $s_{B}^{i}=\left\langle \sigma_{i}\otimes\mathds{1}_{2}\right\rangle $,
$s_{\bar{B}}^{i}=\left\langle \mathds{1}_{2}\otimes\sigma_{i}\right\rangle $
and $C_{ij}=\left\langle \sigma_{i}\otimes\sigma_{j}\right\rangle $.
The Hilbert spaces associated with spin states of $B$ and $\bar{B}$
are denoted as $\mathscr{H}_{B}$ and $\mathscr{H}_{\bar{B}}$ respectively,
thus $\rho_{B\bar{B}}$ describes a quantum state in the joint Hilbert
space $\mathscr{H}_{B}\otimes\mathscr{H}_{\bar{B}}$. The one particle
density operator can be obtained by taking the partial trace $\rho_{B}=\textrm{Tr}_{\bar{B}}(\rho_{B\bar{B}})=\frac{1}{2}(\mathds{1}+\boldsymbol{s}_{B}\cdot\boldsymbol{\sigma})$
and $\rho_{\bar{B}}=\textrm{Tr}_{B}(\rho_{B\bar{B}})=\frac{1}{2}(\mathds{1}+\boldsymbol{s}_{\bar{B}}\cdot\boldsymbol{\sigma})$,
which reduces to Eq.~(\ref{eq:Bloch}) for one qubit. 


\subsection{Joint decay of baryon-antibaryon with spin correlation}

We consider the joint decay $B\bar{B}\rightarrow B^{\prime}M\bar{B}^{\prime}\bar{M}$
with spin correlation. According to the quantum measurement postulate,
a joint decay process can be regarded as \textit{parallel quantum
measurement}. So the joint probability for this parallel measurement
is given by 

\begin{align}
\mathcal{P}(\boldsymbol{n},\bar{\boldsymbol{n}}) & =\textrm{Tr}\left[\Bigl(M_{\boldsymbol{n}}\otimes\bar{M}_{\bar{\boldsymbol{n}}}\Bigr)\rho_{B\bar{B}}\Bigl(M_{\boldsymbol{n}}^{\dagger}\otimes\bar{M}_{\bar{\boldsymbol{n}}}^{\dagger}\Bigr)\right],\label{eq:joint_pr}
\end{align}
similar to Eq.~(\ref{eq:post_Bloch}), where $\boldsymbol{n}$ or
$\bar{\boldsymbol{n}}$ are momentum directions of $B^{\prime}$ and
$\bar{B}^{\prime}$ respectively, and $\{M_{\boldsymbol{n}}\}$ as
well as $\{\bar{M}_{\bar{\boldsymbol{n}}}\}$ are measurement operators
acting on $\mathscr{H}_{B}$ and $\mathscr{H}_{\bar{B}}$ respectively.
In comparison with Eq.~(\ref{eq:angular_re}), the joint probability
is actually the joint angular distribution of the daughter baryon
and antibaryon. By substituting $M_{\boldsymbol{n}}$ and $\bar{M}_{\bar{\boldsymbol{n}}}$
in the form of Eq.~(\ref{eq:M_n}) together with $\rho_{B\bar{B}}$
from Eq.~(\ref{eq:two_qubit}) into Eq.~(\ref{eq:joint_pr}), the
joint angular distribution is given as 
\begin{align}
 & \frac{1}{\Gamma}\frac{d\Gamma(B\bar{B}\rightarrow B^{\prime}M\bar{B}^{\prime}\bar{M})}{d\Omega_{\boldsymbol{n}}d\Omega_{\bar{\boldsymbol{n}}}}\nonumber \\
= & \frac{1}{(4\pi)^{2}}\Bigl(1+\alpha_{B}\boldsymbol{s}_{B}\cdot\boldsymbol{n}+\alpha_{\bar{B}}\boldsymbol{s}_{\bar{B}}\cdot\bar{\boldsymbol{n}}+\alpha_{B}\alpha_{\bar{B}} \sum_{i,j} C_{ij} n_{i}\bar{n}_{j}\Bigr),\label{eq:joint_angular}
\end{align}
where $\alpha_{B}$ and $\alpha_{\bar{B}}$ are decay parameters defined
in Eq.~(\ref{eq:parameters}) associated with $B$ and $\bar{B}$
respectively. 


The similar results can be found in Ref.~\citep{Baumgart:2012ay,Afik:2022dgh}
in the study of the correlated $t\bar{t}$ decay. The joint angular
distribution in Eq.~(\ref{eq:joint_angular}) is derived in the generalized measurement approach. We note that the distribution in Eq.~(\ref{eq:joint_pr})
is different from the one in Eq.~(\ref{eq:decay_chain}), because
the former describes the joint decay of $B\bar{B}$, while the latter
describes the decay chain of $B$. 


\subsection{Charmonium decays and quantum entanglement}

The chamonium $(c\bar{c})$ decays to hyperon-antihyperon provide
an ideal place to test the quantum entanglement and correlation in
the joint decay of hyperon-antihyperon. From Eqs.~(\ref{eq:two_qubit})
and (\ref{eq:joint_angular}) we see that $\rho_{B\bar{B}}$ can be
probed or extracted by the joint angular distribution in the decay
of hyperon-antihyperon in experiments. 


In this work, we focus on the deacys of spin-0 charmonia $\eta_{c}/\chi_{c0}\rightarrow\Lambda\bar{\Lambda}\rightarrow p\pi^{-}\bar{p}\pi^{+}$
which were discussed in Ref.~\citep{Chen:2013epa,Chen:2020pia}. Since $\eta_{c}$
is a pseudoscalar particle, the spin state of $\Lambda\bar{\Lambda}$
should be spin singlet corresponding to the following density operator
\begin{equation}
\rho_{\Lambda\bar{\Lambda}}(\eta_{c})=\left|\Psi^{-}\right\rangle \left\langle \Psi^{-}\right|=\frac{1}{4}\left(\mathds{1}_{4}-\sum_{i}\sigma_{i}\otimes\sigma_{i}\right),
\label{eq:eta}
\end{equation}
where $\left|\Psi^{-}\right\rangle $ denotes spin singlet $\left|\Psi^{-}\right\rangle =(\left|01\right\rangle -\left|10\right\rangle )/\sqrt{2}$.
From Eq.~(\ref{eq:eta}), there is no polarization for $\Lambda$
and $\bar{\Lambda}$ since $\boldsymbol{s}_{\Lambda}=\boldsymbol{s}_{\bar{\Lambda}}=\boldsymbol{0}$,
and the spin correlation matrix reads $C=\textrm{diag}\{-1,-1,-1\}$.
The spin state of $\Lambda\bar{\Lambda}$ in the decay of the scalar
particle $\chi_{c0}$ is the spin-0 state of the spin triplet in the
spin quantization direction, so the spin density operator for $\Lambda\bar{\Lambda}$
reads 
\begin{equation}
\rho_{\Lambda\bar{\Lambda}}(\chi_{c0})=\left|\Psi^{+}\right\rangle \left\langle \Psi^{+}\right|=\frac{1}{4}\left(\mathds{1}_{4}+\sum_{i,j}C_{ij}\sigma_{i}\otimes\sigma_{j}\right),\label{eq:chi}
\end{equation}
where $\left|\Psi^{+}\right\rangle =(\left|01\right\rangle +\left|10\right\rangle )/\sqrt{2}$
is the Bell state in the spin triplet. We see that there is no polarization
for for $\Lambda$ and $\bar{\Lambda}$ and the spin correlation matrix
is $C=\textrm{diag}\{+1,+1,-1\}$. Using Eqs. (\ref{eq:eta}) and
(\ref{eq:chi}) in Eq.~(\ref{eq:joint_angular}), we obtain the joint
angular distributions as 
\begin{align}
\frac{1}{\Gamma}\frac{d\Gamma_{\eta_{c}}}{d\Omega_{\boldsymbol{n}}d\Omega_{\bar{\boldsymbol{n}}}}= & \frac{1}{(4\pi)^{2}}\left(1-\alpha_{\Lambda}\alpha_{\bar{\Lambda}}\boldsymbol{n}\cdot\bar{\boldsymbol{n}}\right),\nonumber \\
\frac{1}{\Gamma}\frac{d\Gamma_{\chi_{c0}}}{d\Omega_{\boldsymbol{n}}d\Omega_{\bar{\boldsymbol{n}}}}= & \frac{1}{(4\pi)^{2}}\left[1+\alpha_{\Lambda}\alpha_{\bar{\Lambda}}(n_{x}\bar{n}_{x}+n_{y}\bar{n}_{y}-n_{z}\bar{n}_{z})\right],\label{eq:char}
\end{align}
where $\boldsymbol{n}$ and $\bar{\boldsymbol{n}}$ are momentum directions
for $p$ and $\bar{p}$ respectively. 


From the above cases, it is evident that the spin density operator
$\rho_{\Lambda\bar{\Lambda}}$ can be reconstructed tomographically
from the joint distribution of $p\bar{p}$ in the subsequent decay
$\Lambda\bar{\Lambda}\rightarrow p\pi^{-}\bar{p}\pi^{+}$. The reconstruction
of $\rho_{\Lambda\bar{\Lambda}}$ enables a comprehensive analysis
of quantum correlation in $\Lambda\bar{\Lambda}$ system. 


Generally speaking, the quantum correlation or entanglement in $\rho_{B\bar{B}}$
is fully encoded in $C_{ij}$. There are various types of quantum
correlation~\citep{Afik:2022dgh}. In this work, we only consider
Bell nonlocality in testing the violation of the CHSH inequality.
The maximum value of the Bell operator associated with the CHSH inequality
can be directly calculated through $C_{ij}$ in $\rho_{B\bar{B}}$.
According to Ref.~\citep{Horodecki:1995nsk}, the maximum value of the
Bell operator reads $\left\langle \mathcal{B}_{\textrm{CHSH}}\right\rangle _{\textrm{max}}=2\sqrt{\lambda_{1}+\lambda_{2}}$,
where $\lambda_{1}$ and $\lambda_{2}$ are two largest eigenvalues
of the matrix $C^{\textrm{T}}C$. Thus, from Eqs.~(\ref{eq:eta})
and (\ref{eq:chi}) in decays of $\eta_{c}$ and $\chi_{c0}$, the
upper bound $2\sqrt{2}$ in the CHSH inequality has been approached
in $\Lambda\bar{\Lambda}$. This violation lies in the fact that the
spin states of $\Lambda\bar{\Lambda}$ are the Bell states $\left|\Psi^{\pm}\right\rangle $
that are maximally entangled. 


\section{Quantum simulation \label{sec:Quantum_simulation}}

In this section, we will perform quantum simulation for the decay of hyperon-antihyperon
and present the test of the CHSH inequality through the spin correlation
in the hyperon-antihyperon system.


\subsection{Simulation for generalized measurement process}

Similar to $M_{\boldsymbol{n}}$ for the hyperon decay in Eq.~(\ref{eq:M_n}),
$\bar{M}_{\boldsymbol{n}}$ for the antihyperon decay is defined as
\begin{equation}
\bar{M}_{\boldsymbol{n}}=\frac{1}{\sqrt{4\pi}}\frac{1}{\sqrt{\left|\bar{S}\right|^{2}+\left|\bar{P}\right|^{2}}}\left(\bar{S}+\bar{P}\boldsymbol{\sigma}\cdot\boldsymbol{n}\right),\label{eq:M_n_bar}
\end{equation}
which can be obtained from Eq.~(\ref{eq:M_n}) by simply making the
replacement $S\to\bar{S}$ and $P\to\bar{P}$. 


In order to perform the simulation on the digital quantum computer,
the measurement operators $\{M_{\boldsymbol{n}}\}$ must be embeded
into unitary operators. It can be verified that the block matrix 
\begin{equation}
U_{\boldsymbol{n}}\equiv\frac{1}{\sqrt{2\left(\left|S\right|^{2}+\left|P\right|^{2}\right)}}\left[\begin{array}{c|c}
S+P\boldsymbol{\sigma}\cdot\boldsymbol{n} & P^{*}-S^{*}\boldsymbol{\sigma}\cdot\boldsymbol{n}\\
\hline S-P\boldsymbol{\sigma}\cdot\boldsymbol{n} & P^{*}+S^{*}\boldsymbol{\sigma}\cdot\boldsymbol{n}
\end{array}\right],\label{eq:U_Mn}
\end{equation}
is unitary, and $M_{\boldsymbol{n}}$ is embeded as the upper-left
block. From the definition of $M_{\boldsymbol{n}}$ in Eq.~(\ref{eq:M_n}),
it can be expressed by $M_{\boldsymbol{n}}=U(\mathcal{R})M_{\hat{\boldsymbol{z}}}U(\mathcal{R})^{\dagger}$,
where $U(\mathcal{R})$ denotes a SU(2) rotation isomorphic to $\mathcal{R}(\phi,\theta,0)$.
As a consequence, the unitary operator $U_{\boldsymbol{n}}$ can be
obtained by performing a similarity transformation on $U_{\hat{\boldsymbol{z}}}$,
as $U_{\boldsymbol{n}}=\left[\mathds{1}\otimes U(\mathcal{R})\right]U_{\hat{\boldsymbol{z}}}\left[\mathds{1}\otimes U(\mathcal{R})^{\dagger}\right]$.
Similarly the unitary operator $\bar{U}_{\boldsymbol{n}}$ for the
antihyperon decay can also be defined by the replacement $S\to\bar{S}$
and $P\to\bar{P}$. 

\begin{figure}[H]
\begin{centering}
\begin{quantikz} 
\lstick{$\ket{0}$} & \qw & \gate[wires = 2][1.2cm]{U_{\hat{\boldsymbol{z}}}} & \qw & \meter{} \\ 
\lstick{$\ket{\Lambda}$} & \gate{U(\mathcal{R})^{\dagger}} & & \gate{U(\mathcal{R})} &  \ \push{\sqrt{2 \pi} M_{\boldsymbol{n}_{p}}\ket{\Lambda}} \\  
\lstick{$\ket{0}$} & \qw & \gate[wires = 2][1.2cm]{\bar{U}_{\hat{\boldsymbol{z}}}} & \qw & \meter{} \\ 
\lstick{$\ket{\bar{\Lambda}}$} & \gate{U(\bar{\mathcal{R}})^{\dagger}} & & \gate{U(\bar{\mathcal{R}})} &  \ \push{\sqrt{2 \pi} \bar{M}_{\boldsymbol{n}_{\bar{p}}}\ket{\bar{\Lambda}}} \\  
\end{quantikz}
\par\end{centering}
\caption{The quantum circuit for generalized measurements. The spin states of $\Lambda$
and $\bar{\Lambda}$ are prepared to be in the Bell states $\left|\Psi^{\pm}\right\rangle $.
The additional two ancilla qubits are initialized to be in $\left|0\right\rangle $.
$U_{\hat{\boldsymbol{z}}}$ and $\bar{U}_{\hat{\boldsymbol{z}}}$are
the unitary gates in which the measurement operators $M_{\hat{\boldsymbol{z}}}$
and $\bar{M}_{\hat{\boldsymbol{z}}}$ are embedded. $U(\mathcal{R})$
and $U(\bar{\mathcal{R}})$ are SU(2) rotation gates. The measurements
are performed on two ancilla qubits by which the quantum information
in $\Lambda\bar{\Lambda}$ can be extracted. \label{fig:char}}
\end{figure}
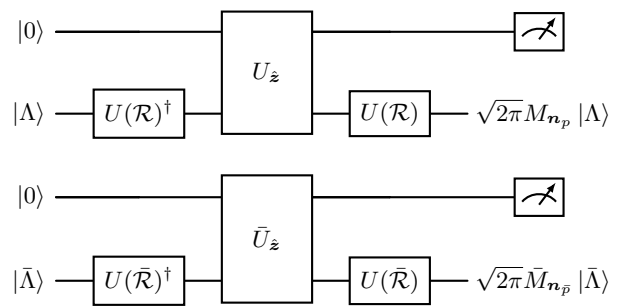

Following this, the quantum circuit for simulating the joint decay
of $\Lambda\bar{\Lambda}\rightarrow p\pi^{-}\bar{p}\pi^{+}$ is presented
in Fig.~\ref{fig:char}, where the spin states of $\Lambda\bar{\Lambda}$
are prepared to be in the Bell states $\left|\Psi^{\pm}\right\rangle $
resulted from $\eta_{c}/\chi_{c0}$ decay as discussed in Sec.~\ref{sec:Joint_decay}.
Figure~\ref{fig:char} indicates that the measurement operators have
been successfully embedded into unitary operators, and the quantum
information contained in $\Lambda\bar{\Lambda}$ can be extracted
from ancilla qubits. 


\begin{figure}[ht]
\begin{centering}
\includegraphics[scale=0.64]{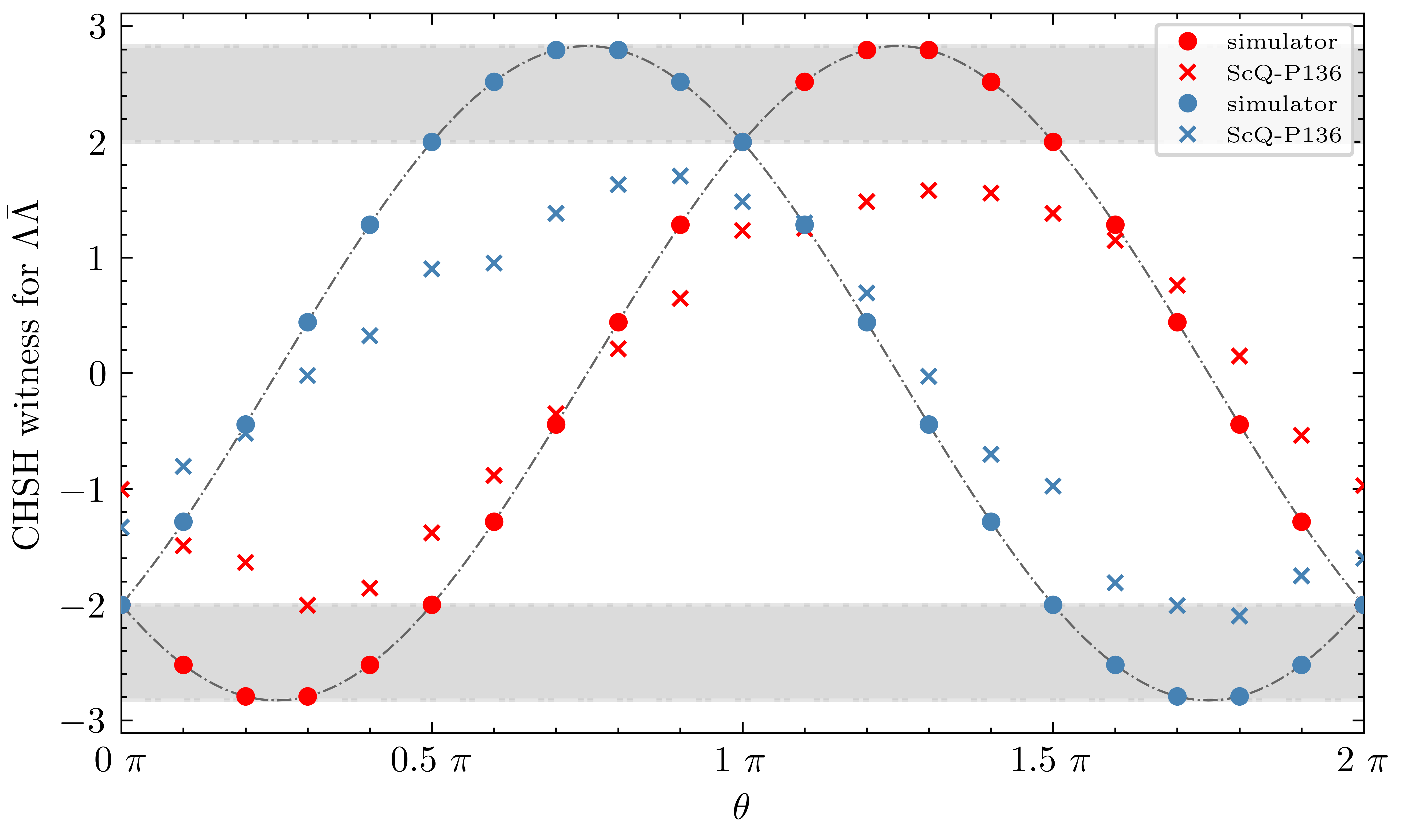}
\par\end{centering}
\caption{The CHSH witness in the simulation of the entangled states of $\Lambda\bar{\Lambda}$.
Solid circles are results on noiseless simulator while cross symbols
are results on the superconducting quantum computer with 136 qubits.
The red/blue symbols represent the initial entangled states $\left|\Psi^{-}\right\rangle /\left|\Psi^{+}\right\rangle $
for $\Lambda\bar{\Lambda}$. The shadow area is the quantum region
$[-2\sqrt{2},-2)\cup(+2+2\sqrt{2}]$. \label{fig:CHSH_Lambda}}
\end{figure}

The simulation has been performed on simulators and the superconducting
quantum computer \emph{Quafu} developed in Beijing Academy of Quantum
Information Sciences. The simulation results are shown in Fig.~\ref{fig:CHSH_Lambda},
which agree with the theoretical expectation that they enter the quantum
region $[-2\sqrt{2},-2)\cup(2,2\sqrt{2}]$ and reach the maximum $2\sqrt{2}$. Thus we conclude that
there is maximum violation of the CHSH inequality in $\Lambda\bar{\Lambda}$
from charmonium decays, which coincides with the result in Sec.~\ref{sec:Joint_decay}.
However, the violation of the CHSH inequality on real quantum computers
turns out to be very weak. This is due to the noise in the real platform,
which decreases the quantumness between two qubits. The details of
our quantum simulation are shown in Appendix.~\ref{sec:sim_de}. 


\subsection{Simulation for depolarizing channel}

As we discussed in Eq.~(\ref{eq:averaged_Bloch}), the average spin
density matrix of the daughter baryon in $B\rightarrow B^{\prime}M$
is $\overline{\rho}_{B^{\prime}}=\mathcal{E}(\rho_{B})$. In the viewpoint
of QIS, the post-measurement ensemble can be interpreted as a quantum
evolution characterized by the quantum channel. We notice that Eq.~(\ref{eq:averaged_Bloch})
can be regarded as a single-qubit depolarizing channel as 
\begin{equation}
\rho_{B}\mapsto\rho_{B^{\prime}}\equiv\mathcal{E}(\rho_{B})= \mathscr{P} \frac{\mathds{1}}{2}+(1-\mathscr{P})\rho_{B},
\end{equation}
where $\mathscr{P}\equiv(2-2\gamma_{B})/3$. We have a similar form
for antibaryon: $\rho_{\bar{B}^{\prime}}\equiv\bar{\mathcal{E}}(\rho_{\bar{B}})$
with $\bar{\mathscr{P}}\equiv(2-2\gamma_{\bar{B}})/3$. Then the average
spin density matrix for $B^{\prime}\bar{B}^{\prime}$ from the joint
decay of $B\bar{B}\rightarrow B^{\prime}M\bar{B}^{\prime}\bar{M}$
can be presented as 
\begin{eqnarray}
\rho_{B^{\prime}\bar{B}^{\prime}}  & = & \mathcal{E}\otimes\bar{\mathcal{E}}(\rho_{B\bar{B}}) \nonumber \\
 & = & \frac{1}{4}\biggl[\mathds{1}+(1-\mathscr{P})\boldsymbol{\sigma}\cdot\boldsymbol{s}_{B}\otimes\mathds{1}+(1-\bar{\mathscr{P}})\mathds{1}\otimes\boldsymbol{\sigma}\cdot\boldsymbol{s}_{\bar{B}} \nonumber \\
 & &  +(1-\mathscr{P})(1-\bar{\mathscr{P}})\sum_{i,j}C_{ij}\sigma_{i}\otimes\sigma_{j}\biggr].
\label{eq:depolarization}
\end{eqnarray}
Since the decay processes of $B$ and $\bar{B}$ are not correlated,
the two-qubit channel can be described as a tensor product of two
independent single-qubit depolarizing channels $\mathcal{E}\otimes\bar{\mathcal{E}}$.
From Eq.~(\ref{eq:depolarization}), the polarization vectors decrease
by factors $(1-\mathscr{P})$ and $(1-\bar{\mathscr{P}})$, while
the spin correlation decreases by a factor $(1-\mathscr{P})(1-\bar{\mathscr{P}})$.
This means that the spin correlation is suppressed in decay processes.
The suppression of the spin correlation may lead to the satisfaction
of the CHSH inequality, since the maximal violation also decreases
by the factor $(1-\mathscr{P})(1-\bar{\mathscr{P}})$, which has been
discussed in Ref.~\citep{Qian:2020ini}. 

\begin{figure}[ht]
\begin{centering}
\includegraphics[scale=0.64]{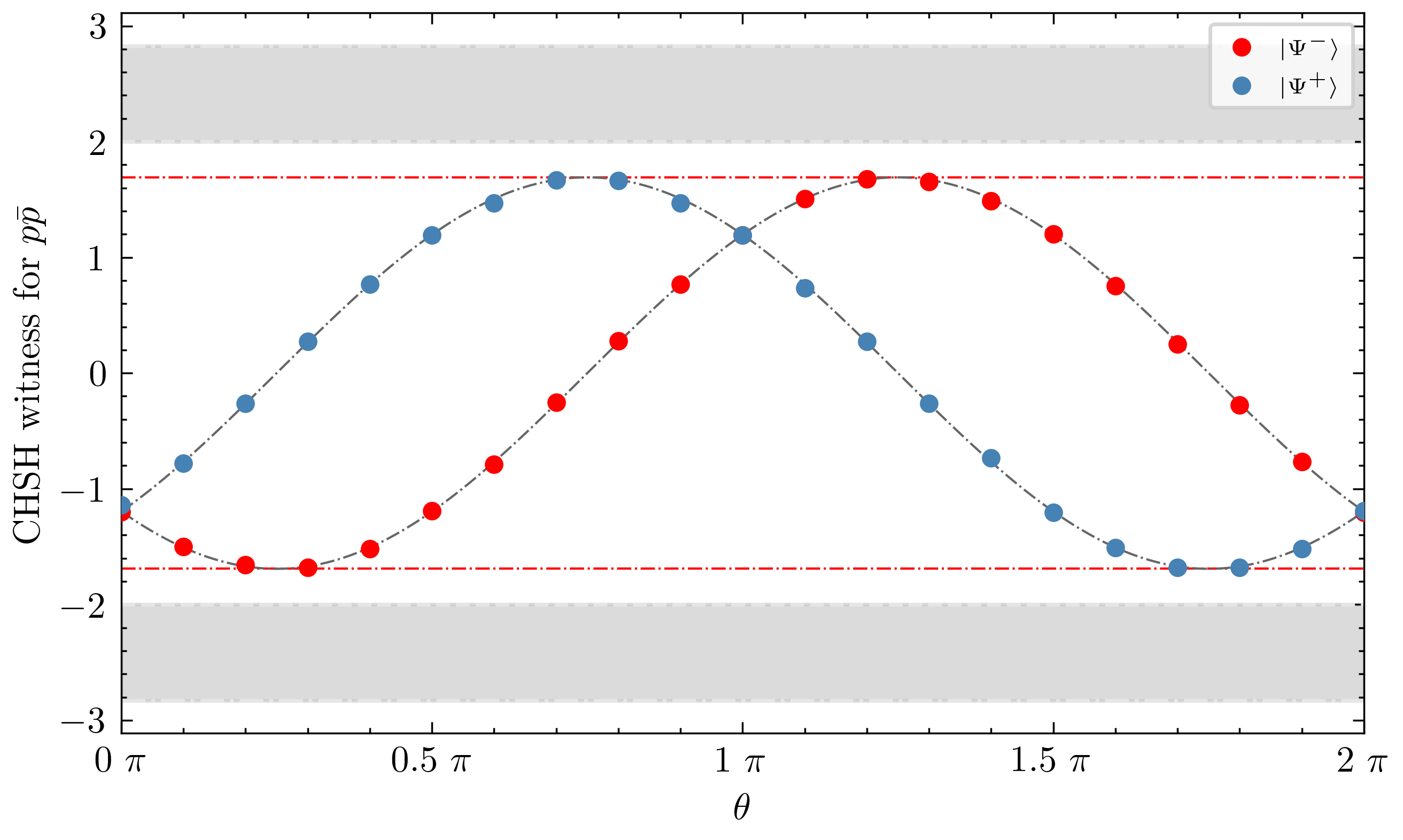}
\par\end{centering}
\caption{The simulation results for $p\bar{p}$ in the decay $\eta_{c}/\chi_{c0}\rightarrow\Lambda\bar{\Lambda}\rightarrow p\pi^{-}\bar{p}\pi^{+}$
and test of the CHSH inequality. The solid circles denote the results
on the noiseless simulator. The red/blue symbols represent $p\bar{p}$
produced in the subsequent decay of $\Lambda\bar{\Lambda}$ in prepared
initial states $\left|\Psi^{-}\right\rangle /\left|\Psi^{+}\right\rangle $.
The shadow area is the quantum region of $[-2\sqrt{2},-2)\cup(+2+2\sqrt{2}]$,
and the red dashdotted lines indicate the bound in Eq.~(\ref{eq:CHSH}).
\label{fig:CHSH_p}}
\end{figure}

We now perform a simulation based on decays $\eta_{c}/\chi_{c0}\rightarrow\Lambda\bar{\Lambda}\rightarrow p\pi^{-}\bar{p}\pi^{+}$. The simulation results for the channel $\rho_{p\bar{p}}=\mathcal{E}\otimes\bar{\mathcal{E}}(\rho_{\Lambda \bar{\Lambda}})$
are presented in Fig.~\ref{fig:CHSH_p}. In principle, the maximum
value of the Bell operator in $p\bar{p}$ becomes 
\begin{equation}
\left\langle \mathcal{B}_{\textrm{CHSH}}\right\rangle _{\textrm{max}}=\left(\frac{1+2\gamma_{\Lambda}}{3}\right)^{2}2\sqrt{2},
\label{eq:CHSH}
\end{equation}
with $\gamma_{\Lambda} = \gamma_{\bar{\Lambda}}$.  The data for $\Lambda$ hyperon gives $\gamma_{\Lambda}\approx0.66$
in PDG~\citep{ParticleDataGroup:2020ssz}. Thus, the maximum value is $\left\langle \mathcal{B}_{\textrm{CHSH}}\right\rangle _{\textrm{max}}\approx1.69<2$,
which means the spin correlation of $p\bar{p}$ cannot not reach the
quantum bound and does not show the property of nonlocality.


In Fig.~\ref{fig:CHSH_p}, we can observe that the simulation results
are in agreement with our theoretical expectation that the spin correlation
in decay daughters decreases in decay processes. 


\section{Projective measurement and unitary evolution \label{sec:Projective_me} }

In this section, we will discuss the origin of generalized measurement
and its connection with particle decay/scattering processes. 

In the hyperon decay $B\to B^{\prime}M$, since $M$ is a pseudoscalar
meson, we can neglect its quantum state. The quantum state of the
mother hyperon is $\left|B\right\rangle =\left|\textrm{momentum}\right\rangle \otimes\left|\textrm{spin}\right\rangle $
in the Hilbert space $\mathscr{H}_{\textrm{momentum}}\otimes\mathscr{H}_{\textrm{spin}}$.
The general decay process can be described by a unitary evolution
$U$ called the scattering matrix $S$ in quantum field theory. In
the decay, the outgoing daughter baryon is detected in the direction
$\boldsymbol{n}(\theta,\phi)$ with the angular distribution in experiments.


From the perspective of QIS, the decay process can be described as
a unitary evolution governed by the Hamiltonian (or Lagrangian) of
the system, $U(\rho_{\textrm{momentum}}\otimes\rho_{\textrm{spin}})U^{\dagger}$.
Then the decay takes place by an emission of daughter particles in
a certain direction, which indicates that a projective measurement
is performed on the evolved system as $\Pi_{\theta\phi}U(\rho_{\textrm{momentum}}\otimes\rho_{\textrm{spin}})U^{\dagger}\Pi_{\theta\phi}^{\dagger}$,
where the measurement operators are defined as projectors $\Pi_{\theta\phi}=\left|\theta\phi\right\rangle \left\langle \theta\phi\right|\in\mathscr{H}_{\textrm{momentum}}$.
Here the quantum state of the daughter baryon $\rho_{B^{\prime}}=U(\rho_{\textrm{momentum}}\otimes\rho_{\textrm{spin}})U^{\dagger}$
has been projected into the direction $\boldsymbol{n}(\theta,\phi)$
in momentum space. Note that the momentum magnitude $|\boldsymbol{p}|$
is fixed by the energy-momentum conservation in the two-body decay. 


According to the quantum measurement postulate discussed in Sec.~\ref{subsec:general_mea},
the probability for finding the daughter baryon in the $\boldsymbol{n}$
direction reads 
\begin{equation}
\mathcal{P}(\boldsymbol{n})\propto\textrm{Tr}\left(\Pi_{\theta\phi}U(\rho_{\textrm{momentum}}\otimes\rho_{\textrm{spin}})U^{\dagger}\Pi_{\theta\phi}^{\dagger}\right),\label{eq:projective_me}
\end{equation}
where $\mathcal{P}(\boldsymbol{n})$ is just the angular distribution
$(1/\Gamma)(d\Gamma/d\Omega)$. In decay or scattering processes,
the momenta of initial particles are pre-determined. In the decay
process, the initial momentum state is denoted as $\left|\boldsymbol{0}\right\rangle =\left|\boldsymbol{p}=\boldsymbol{0}\right\rangle $
by choosing the rest frame of mother particle, hence we have $\rho_{\textrm{momentum}}^{B}=\left|\boldsymbol{0}\right\rangle \left\langle \boldsymbol{0}\right|$
in Eq.~(\ref{eq:projective_me}). Since the initial momentum state
is fixed, and the density matrix only contains the spin degree of
freedom for the daughter baryon and can be obtained by taking the
partial trace over momentum
\begin{equation}
\rho_{\textrm{spin}}^{B^{\prime}}(\boldsymbol{n})\propto\textrm{Tr}_{\textrm{momentum}}\left(\Pi_{\theta\phi}U\left|\boldsymbol{0}\right\rangle \rho_{\textrm{spin}}^{B}\left\langle \boldsymbol{0}\right|U^{\dagger}\Pi_{\theta\phi}^{\dagger}\right),\label{eq:spin_pro}
\end{equation}
where the action $\textrm{Tr}_{\textrm{momentum}}$ discards the momentum
degree of freedom. Considering Eqs.~(\ref{eq:post}) and (\ref{eq:post_Bloch}),
it can be shown that $\rho_{\textrm{spin}}^{B^{\prime}}(\boldsymbol{n})\propto M_{\boldsymbol{n}}\rho_{\textrm{spin}}^{B}M_{\boldsymbol{n}}^{\dagger}$,
where $\{M_{\boldsymbol{n}}\}$ denotes the generalized measurement operators
induced by the partial trace over momentum. 


A heuristic quantum circuit in demonstrating the decay $B\to B^{\prime}M$
is shown as 
\begin{center}
\begin{quantikz} 
\lstick{$\ket{\boldsymbol{0}}_{\textrm{momentum}}^{B}$} & \gate[wires = 2][1.5cm]{U} & \meter{\Pi_{\theta\phi}} & \cw \rstick{$\mathcal{P}(\boldsymbol{n})$} \\ 
\lstick{$\rho_{\textrm{spin}}^{B}$}  &  &  \qw \rstick{$\rho_{\textrm{spin}}^{B^{\prime}}(\boldsymbol{n})$}  
\end{quantikz}
\par\end{center}
This quantum circuit gives a pedagogical illustration of the projective
measurement and unitary evolution in particle decay and scattering
processes in accordance with Eqs.~(\ref{eq:projective_me}, \ref{eq:spin_pro}).
The upper wire denotes the momentum state of the mother baryon $\left|\boldsymbol{p}\right\rangle $,
which is initialized as $\left|\boldsymbol{0}\right\rangle $ in the
rest frame, while the lower one denotes the spin state. The unitary
gate ``$U$'' represents the unitary S-matrix in scattering theory.
After the unitary evolution, a projective measurement $\Pi_{\theta\phi}$
is performed on the momentum state, resulting in the angular distribution
of the daughter baryon. Consequently, the spin density operator $\rho_{\textrm{spin}}^{B^{\prime}}(\boldsymbol{n})$
in Eq.~(\ref{eq:spin_pro}) is then obtained after the projective
measurement. 


In quantum information theory, a unitary evolution combined with projective
measurements are suffice to induce a set of generalized measurements $\{M_{\boldsymbol{n}}\}$
which contain both the dynamical and measurement information of the
system. The generalized measurement formalism is very useful to describe
decay and scattering processes in particle physics. For more details
of the topic, we refer the readers to Chapter 2.2.8 of Ref.~\citep{Nielsen2015}. 


\section{Summary \label{sec:discussions}}

The particle decay processes is described as the generalized quantum measurement
in quantum information theory. We consider a spin-1/2 hyperon decaying
to one spin-1/2 baryon and one spin-0 meson. In this perspective,
we successfully establish a correspondence between the angular distribution
of the daughter baryon and the generalized measurement operator. The Bloch-Fano
representation is employed to describe the quantum measurement process,
which shows a direct parallelism with the decay matrices outlined
in Ref.~\citep{Batozskaya:2023rek}. We apply this method to the joint
decay of $\eta_{c}/\chi_{c0}\rightarrow\Lambda\bar{\Lambda}\to p\pi^{+}\bar{p}\pi^{-}$
and investigate the spin correlation in $\Lambda\bar{\Lambda}$ as
well as in $p\bar{p}$ systems. The quantum simulation to test the
CHSH inequality on both the simulator and real quantum computer has
been done. A discussion on the connection between the particle decay
and the generalized quantum measurement has been given. 


The generalized measurement description can be applied to a wide range
of decay and scattering processes. In particular, it offers us a QIS-based
tool to analyze unknown particle decays in search for new physics,
such as dark matter particles. It also opens a window for testing
quantum nonlocality or other quantum properties in particle decays
on quantum computers. Our results can be verified in particle experiments
such as BESIII~\citep{BESIII:2018cnd}.

\begin{acknowledgments}
We would like to thank S. Lin, D. E. Liu and R. Venugopalan for helpful discussions. 
And we thank H.-Z. Xu for \emph{Quafu} software and hardware supports. 
This work is supported by the National Natural Science Foundation of China (NSFC) under Grant Nos. 12135011, 12305010, and by the Strategic Priority Research Program of the Chinese Academy of Sciences (CAS) under Grant No. XDB34030102. 
\end{acknowledgments}

\appendix

\section{Details for quantum simulation \label{sec:sim_de}}

Through the unitary embedding introduced in Sec.~\ref{sec:Quantum_simulation},
we have the unitary operators $U_{\boldsymbol{n}}$ and $\bar{U}_{\boldsymbol{n}}$
acting on the corresponding states as 
\begin{align}
U_{\boldsymbol{n}}:\left|0\right\rangle \otimes & \left|\Lambda\right\rangle \mapsto \underset{\Lambda\rightarrow p}{\underbrace{\sqrt{2\pi}\left|0\right\rangle \otimes M_{\boldsymbol{n}}\left|\Lambda\right\rangle }}\nonumber \\
 & +\frac{1}{\sqrt{2\left(\left|S\right|^{2}+\left|P\right|^{2}\right)}}\left|1\right\rangle \otimes\left(S-P\boldsymbol{\sigma}\cdot\boldsymbol{n}\right)\left|\Lambda\right\rangle ,\nonumber \\
\bar{U}_{\boldsymbol{n}}:\left|0\right\rangle \otimes & \left|\bar{\Lambda}\right\rangle  \mapsto \underset{\bar{\Lambda}\rightarrow\bar{p}}{\underbrace{\sqrt{2\pi}\left|0\right\rangle \otimes\bar{M}_{\boldsymbol{n}}\left|\bar{\Lambda}\right\rangle }}\nonumber \\
 & +\frac{1}{\sqrt{2\left(\left|\bar{S}\right|^{2}+\left|\bar{P}\right|^{2}\right)}}\left|1\right\rangle \otimes\left(\bar{S}-\bar{P}\boldsymbol{\sigma}\cdot\boldsymbol{n}\right)\left|\bar{\Lambda}\right\rangle ,\label{eq:unitary}
\end{align}
where $\left|0\right\rangle =(1,0)^{\mathrm{T}}$, $\left|1\right\rangle =(0,1)^{\mathrm{T}}$,
$\left|0\right\rangle \otimes\left|\Lambda\right\rangle =(\lambda_{1},\lambda_{2},0,0)^{\mathrm{T}}$
and $\left|1\right\rangle \otimes\left|\Lambda\right\rangle =(0,0,\lambda_{1},\lambda_{2})^{\mathrm{T}}$
with $\left|\Lambda\right\rangle =(\lambda_{1},\lambda_{2})^{\mathrm{T}}$,
$M_{\boldsymbol{n}}$ is given in Eq.~(\ref{eq:M_n}), and $\bar{M}_{\boldsymbol{n}}$
is given in Eq.~(\ref{eq:M_n_bar}) which are the measurement operators
associated with the decay of $\Lambda$ and $\bar{\Lambda}$ respectively.
From Eq.~(\ref{eq:unitary}), performing the projective measurements
on the ancilla registers results in
\begin{align}
\mathcal{P}(\left|0\right\rangle ) & =(2\pi)\left\langle \Lambda\right|M_{\boldsymbol{n}}^{\dagger}M_{\boldsymbol{n}}\left|\Lambda\right\rangle =\frac{1}{2}\left\langle \Lambda\right|(1+\alpha\boldsymbol{\sigma}\cdot\boldsymbol{n})\left|\Lambda\right\rangle ,\nonumber \\
\mathcal{P}(\left|1\right\rangle ) & =\frac{\left\langle \Lambda\right|\left(S^{*}-P^{*}\boldsymbol{\sigma}\cdot\boldsymbol{n}\right)\left(S-P\boldsymbol{\sigma}\cdot\boldsymbol{n}\right)\left|\Lambda\right\rangle}{2\left(\left|S\right|^{2}+\left|P\right|^{2}\right)} \nonumber \\
 & =\frac{1}{2}\left\langle \Lambda\right|(1-\alpha\boldsymbol{\sigma}\cdot\boldsymbol{n})\left|\Lambda\right\rangle ,
\end{align}
which leads to the expectation value of $\sigma_{z}$ on the ancilla
qubit as 
\begin{equation}
\left\langle \sigma_{z}\right\rangle _{\mathrm{anc}}=\mathcal{P}(\left|0\right\rangle )-\mathcal{P}(\left|1\right\rangle )=\alpha\left\langle \Lambda\right|\boldsymbol{\sigma}\cdot\boldsymbol{n}\left|\Lambda\right\rangle =\alpha\left\langle \boldsymbol{\sigma}\cdot\boldsymbol{n}\right\rangle _{\Lambda}.\label{eq:exp}
\end{equation}
We see in the above formula that the expectation value associated
on the $\Lambda$ qubit can be obtained from the expectation value
of $\sigma_{z}$ on the ancilla qubit $\left\langle \boldsymbol{\sigma}\cdot\boldsymbol{n}\right\rangle _{\Lambda}=(1/\alpha)\left\langle \sigma_{z}\right\rangle _{\mathrm{anc}}$.
The same result holds for $\bar{\Lambda}$. Thus, the joint expectation
value of $\Lambda\bar{\Lambda}$ reads 
\begin{equation}
\left\langle \boldsymbol{\sigma}\cdot\boldsymbol{n}_{p}\otimes\boldsymbol{\sigma}\cdot\boldsymbol{n}_{\bar{p}}\right\rangle _{\Lambda\bar{\Lambda}}\equiv\frac{1}{\alpha\bar{\alpha}}\left\langle \sigma_{z}\otimes\sigma_{z}\right\rangle _{\mathrm{anc}},\label{eq:jiont_exp}
\end{equation}
which can be directly implemented on the quantum circuit and is adopted
in our test of the CHSH inequality on the quantum computer. In our
simulations, the $\mathcal{CP}$ violation is ignored and the parameters
are specified as $\alpha_{\Lambda}=-\alpha_{\bar{\Lambda}}=0.75$
and $\phi_{\Lambda}=\phi_{\bar{\Lambda}}=0$. 

\begin{figure}[ht]
\begin{centering}
\includegraphics[scale=0.35]{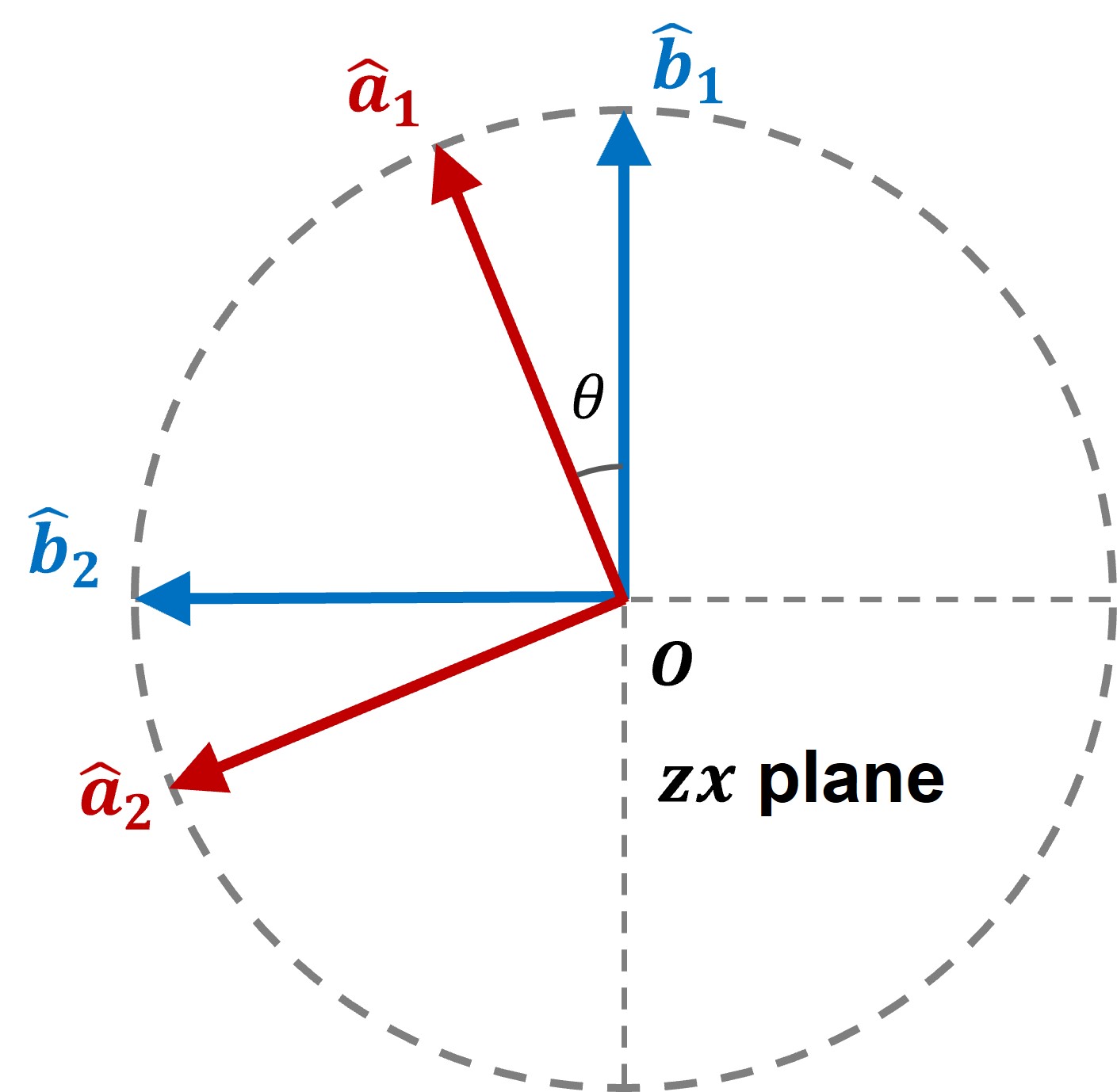}
\par\end{centering}
\caption{Orientations of four unit vectors $\hat{\boldsymbol{a}}_{1}$, $\hat{\boldsymbol{a}}_{2}$,
$\hat{\boldsymbol{b}}_{1}$ and $\hat{\boldsymbol{b}}_{2}$ in $zx$
plane. \label{fig:orientation}}
\end{figure}

The Bell operator defined in Ref.~\citep{Horodecki:1995nsk} reads
\begin{align}
\bigl\langle\mathcal{B}_{\textrm{CHSH}}\bigr\rangle= & \bigl\langle\boldsymbol{\sigma}\cdot\hat{\boldsymbol{a}}_{1}\otimes\boldsymbol{\sigma}\cdot\hat{\boldsymbol{b}}_{1}\bigr\rangle+\bigl\langle\boldsymbol{\sigma}\cdot\hat{\boldsymbol{a}}_{2}\otimes\boldsymbol{\sigma}\cdot\hat{\boldsymbol{b}}_{1}\bigr\rangle\nonumber \\
 & +\bigl\langle\boldsymbol{\sigma}\cdot\hat{\boldsymbol{a}}_{1}\otimes\boldsymbol{\sigma}\cdot\hat{\boldsymbol{b}}_{2}\bigr\rangle-\bigl\langle\boldsymbol{\sigma}\cdot\hat{\boldsymbol{a}}_{2}\otimes\boldsymbol{\sigma}\cdot\hat{\boldsymbol{b}}_{2}\bigr\rangle,\label{eq:Bell_op}
\end{align}
where $\hat{\boldsymbol{a}}_{1}$, $\hat{\boldsymbol{a}}_{2}$, $\hat{\boldsymbol{b}}_{1}$
and $\hat{\boldsymbol{b}}_{2}$ are unit vectors and can be put in
the $zx$ plane. In our paper, the quantum states used in the test
of the CHSH inequality are two Bell states $\left|\Psi^{\pm}\right\rangle $.
For convenience, the Bell operator in Eq.~(\ref{eq:Bell_op}) is
modified as
\begin{align}
\bigl\langle\mathcal{B}_{\textrm{CHSH}}\bigr\rangle_{\Psi^{-}}= & \bigl\langle\hat{\boldsymbol{a}}_{1},\hat{\boldsymbol{b}}_{1}\bigr\rangle+\bigl\langle\hat{\boldsymbol{a}}_{1},\hat{\boldsymbol{b}}_{2}\bigr\rangle-\bigl\langle\hat{\boldsymbol{a}}_{2},\hat{\boldsymbol{b}}_{1}\bigr\rangle+\bigl\langle\hat{\boldsymbol{a}}_{2},\hat{\boldsymbol{b}}_{2}\bigr\rangle,\nonumber \\
\bigl\langle\mathcal{B}_{\textrm{CHSH}}\bigr\rangle_{\Psi^{+}}= & \bigl\langle\hat{\boldsymbol{a}}_{1},\hat{\boldsymbol{b}}_{1}\bigr\rangle+\bigl\langle\hat{\boldsymbol{a}}_{1},\hat{\boldsymbol{b}}_{2}\bigr\rangle+\bigl\langle\hat{\boldsymbol{a}}_{2},\hat{\boldsymbol{b}}_{1}\bigr\rangle-\bigl\langle\hat{\boldsymbol{a}}_{2},\hat{\boldsymbol{b}}_{2}\bigr\rangle,
\end{align}
where $\bigl\langle\hat{\boldsymbol{a}},\hat{\boldsymbol{b}}\bigr\rangle\equiv\bigl\langle\boldsymbol{\sigma}\cdot\hat{\boldsymbol{a}}\otimes\boldsymbol{\sigma}\cdot\hat{\boldsymbol{b}}\bigr\rangle$.
For further simplification, we set the angle between $\hat{\boldsymbol{a}}_{1}$
and $\hat{\boldsymbol{a}}_{2}$ is equal to $\pi/2$, so is the angle
between $\hat{\boldsymbol{b}}_{1}$ and $\hat{\boldsymbol{b}}_{2}$.
Without losing generality, we assume $\hat{\boldsymbol{b}}_{1}=\hat{\boldsymbol{z}}$,
$\hat{\boldsymbol{b}}_{2}=\hat{\boldsymbol{x}}$, $\hat{\boldsymbol{a}}_{1}=\cos\theta\hat{\boldsymbol{z}}+\sin\theta\hat{\boldsymbol{x}}$,
and $\hat{\boldsymbol{a}}_{2}=-\sin\theta\hat{\boldsymbol{z}}+\cos\theta\hat{\boldsymbol{x}}$,
see Fig.~\ref{fig:orientation}. The expectation values are calculated
using the property $\langle\hat{\boldsymbol{a}},\hat{\boldsymbol{b}}\rangle=(\hat{\boldsymbol{a}},C\hat{\boldsymbol{b}})=\hat{\boldsymbol{a}}^{\textrm{T}}C\hat{\boldsymbol{b}}$
with $C$ being the correlation matrix in Eq\@.~(\ref{eq:two_qubit}).
For $\left|\Psi^{-}\right\rangle $ we have $C=\textrm{diag}\{-1,-1,-1\}$,
while for $\left|\Psi^{+}\right\rangle $ we have $C=\textrm{diag}\{+1,+1,-1\}$.
Finally we obtain the results for $\left\langle \mathcal{B}_{\textrm{CHSH}}\right\rangle $
as 
\begin{align}
\left\langle \mathcal{B}_{\textrm{CHSH}}\right\rangle _{\Psi^{-}}= & -2\sqrt{2}\sin(\theta+\frac{\pi}{4}),\nonumber \\
\left\langle \mathcal{B}_{\textrm{CHSH}}\right\rangle _{\Psi^{+}}= & 2\sqrt{2}\sin(\theta-\frac{\pi}{4}).\label{eq:CHSH_re}
\end{align}

\begin{figure}[ht]
\begin{centering}
\begin{quantikz} 
\lstick{$\ket{\Lambda}$} & \gate{H} & \ctrl{1} & \gate{\mathcal{E}} \gategroup[2,steps=1,style={dashed, rounded corners,fill=blue!20,inner sep=10pt},background, ,label style={label position=below,anchor=north,yshift=-0.2cm}]{\textrm{depolarozing channel}} & \qw & \ \push{\ket{p}} \\ 
\lstick{$\ket{\bar{\Lambda}}$} & \qw & \targ{} &  \gate{\bar{\mathcal{E}}} & \qw & \ \push{\ket{\bar{p}}} 
\end{quantikz}
\par\end{centering}
\caption{The quantum circuit simulating $\Lambda\bar{\Lambda}\rightarrow p\bar{p}$,
where $\mathcal{E}\otimes\bar{\mathcal{E}}$ represents two depolarizing
channels. \label{fig:channel}}
\end{figure}
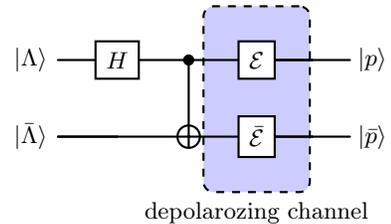

\begin{figure}[H]
\begin{centering}
\begin{quantikz} 
\lstick{$\ket{0}$} & \gate{R_{y}(\varphi)} & \ctrl{3} & \qw & \qw & \qw  \rstick[wires=3]{Trace out}  \\ 
\lstick{$\ket{0}$} & \gate{R_{y}(\varphi)} & \qw & \ctrl{2} & \qw & \qw \\ 
\lstick{$\ket{0}$} & \gate{R_{y}(\varphi)} & \qw & \qw & \ctrl{1} & \qw \\  
\lstick{$\rho$} & \qw & \gate{X} & \gate{Y} & \gate{Z} & \qw \rstick{$\mathcal{E}(\rho)$} 
\end{quantikz}
\par\end{centering}
\caption{Circuit implementation of depolarizing channel. The first three qubits
are introduced as ancillae, while the last is the system qubit representing
the spin of $\Lambda$ hyperon. $X$, $Y$, and $Z$ are three Pauli
gates and $\varphi$ in $R_{y}$ rotation gates are $\varphi=\frac{1}{2}\arccos(1-2\mathscr{P})$,
with $\mathscr{P}$ in Eq.~(\ref{eq:depolar}). \label{fig:implementation}}
\end{figure}
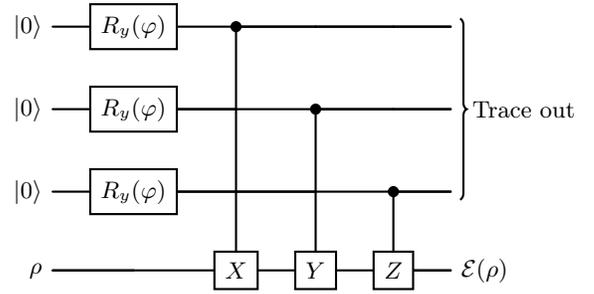

In the simulation of the spin correlation in $p\bar{p}$ in the jointed
decay of $\Lambda\bar{\Lambda}\rightarrow p\pi^{-}\bar{p}\pi^{+}$,
we implement the depolarizing channel by the quantum circuit in Fig.~(\ref{fig:channel}).
The single-qubit depolarizing channel can be expressed as 
\begin{equation}
\mathcal{E}(\rho)=\left(1-\frac{3\mathscr{P}}{4}\right)\rho+\frac{\mathscr{P}}{4}\left(X\rho X+Y\rho Y+Z\rho Z\right),\label{eq:depolar}
\end{equation}
and the details for the quantum circuit implementation are shown in
Fig.~(\ref{fig:implementation}). 


\bibliographystyle{apsrev4-2}
\phantomsection
\addcontentsline{toc}{section} {\refname}
\bibliography{refs}

\end{document}